\documentclass[12pt]{amsart}
\usepackage{amssymb}
\begin{document}
\def\version{File S02v5b.tex last changed 4 April 2002 PBG}
\def\nmonth{\ifcase\month\ \or January\or
   February\or March\or 
April\or May\or June\or July\or August\or
   September\or October\or
November\else December\fi}
\def\nmonth{\ifcase\month\ \or January\or
   February\or March\or April\or May\or June\or July\or August\or
   September\or October\or November\else December\fi}
\def\rightheadline{\hfill\folio\hfill}
\def\leftheadline{\hfill\folio\hfill} 
\newtheorem{theorem}{Theorem}[section]
\newtheorem{lemma}[theorem]{Lemma}
\newtheorem{remark}[theorem]{Remark}
\def\operatorname#1{{\rm#1\,}}
\def\text#1{{\hbox{#1}}}
\def\qedbox{\hbox{$\rlap{$\sqcap$}\sqcup$}}
\def\BB{{\mathcal{B}_1}}
\def\CC{{\mathcal{B}_2}}
\def\BX{{\mathcal{B}_{DR}}}
\def\BD{{\mathcal{B}_D}}
\def\BR{{\mathcal{B}_R}}
\def\B{{\mathcal{B}}}
\def\tr{{\operatorname{Tr}}}
\def\dvol{{\operatorname{dvol}}}
\makeatletter
  \renewcommand{\theequation}{%
   \thesection.\arabic{equation}}
  \@addtoreset{equation}{section}
 \makeatother
\def\coea{a_1}
\def\coeb{a_2}
\def\coec{a_3}
\def\coed{a_4}
\def\coee{a_5}
\def\coef{a_6}
\def\coeg{a_7}
\def\coeh{a_8}
\def\coei{a_9}
\def\coej{a_{10}}
\def\coek{a_{11}}
\def\coel{a_{12}}
\def\cfa{a_{20}}
\def\cfb{a_{21}}
\def\cfc{a_{22}}
\def\cfd{a_{23}}
\def\cfe{a_{24}}
\def\cff{a_{25}}
\def\cfg{a_{26}}
\def\cfh{a_{27}}
\def\cfi{a_{28}}
\def\cfj{a_{29}}
\def\cfk{a_{30}}
\def\cfl{a_{31}}
\def\cfm{a_{32}}
\def\cfmx{a_{33}}
\def\cfn{a_{34}}
\def\cfo{a_{35}}
\def\cfp{a_{36}}
\def\cfq{a_{37}}
\def\cfr{a_{38}}
\def\cfs{a_{39}}
\def\cft{a_{40}}
\def\cfu{a_{41}}
\def\cfv{a_{42}}
\def\cfw{a_{43}}
\def\cfx{a_{44}}
\def\cfy{a_{45}}
\def\cfz{a_{46}}
\def\cfaa{a_{47}}
\def\cfab{a_{48}}
\def\cfac{a_{49}}
\def\cfad{a_{50}}
\def\cfae{a_{51}}
\def\cfaf{a_{52}}
\def\cfag{a_{53}}
\def\cfah{a_{54}}
\def\cfai{a_{55}}
\def\cfaj{a_{56}}
\def\cfak{a_{57}}
\def\cfal{a_{58}}
\def\cfam{a_{59}}
\def\cfan{a_{60}}
\def\doea{b_1}
\def\doeb{b_2}
\def\doec{b_3}
\def\doed{b_4}
\def\doee{b_5}
\def\doef{b_6}
\def\doeg{b_7}%%Last one used
\def\dfa{b_{10}}
\def\dfb{b_{11}}
\def\dfc{b_{12}}
\def\dfd{b_{13}}
\def\dfe{b_{14}}
\def\dff{b_{15}}
\def\dfg{b_{16}}
\title[Heat content asymptotics]{Heat content asymptotics with transmittal and transmission boundary conditions}
\author{P. Gilkey
%\thanks{
%Research partially supported by the NSF (USA) and MPI 
%(Leipzig)}
and K. Kirsten
%\thanks{Research partially supported by the MPI (Leipzig)}
}\begin{abstract}
We study the heat content asymptotics on a Riemannian manifold with smoooth
boundary defined by Dirichlet,  Neumann, transmittal and
transmission boundary conditions. \newline Subject Classification: 
58J50
\end{abstract}
\maketitle

\section{Introduction}\label{Sect1} Let $M$ be a compact $m$ dimensional
Riemannian manifold
 with smooth boundary $\partial M$. Let
$D$ be an operator of Laplace  type on a vector bundle $V$ over $M$. Let 
$\B$ be a suitable local boundary condition and let $D_\B$ be the associated
realization. Let
$\phi\in C^\infty(V)$ describe the initial temperature distribution. 
The subsequent temperature distribution $u:=e^{-tD_\B}\phi$ for
$t\ge0$ is described by the equations:
\begin{equation}(\partial_t+D)u=0,\ u(x;0)=\phi,\text{ and }\B 
u=0.\label{AREFaa}\end{equation}

The specific heat $\rho$ is a section to the dual bundle $V^*$. Let
$$\beta(\phi,\rho,D,\B)(t):=\textstyle\int_Mu\rho$$
be the total heat energy content. As $t\downarrow 0$, there is a complete
asymptotic expansion of the  form
$$\beta(\phi,\rho,D,\B)(t)\sim\textstyle\sum_{n\ge0}\beta_n(\phi,\rho,D,\B)t^{n/2};$$
the {\it heat content coefficients} $\beta_n(\phi,\rho,D,\B)$ are 
locally computable. 

If $D_\B$ is self-adjoint, then let $\{\phi_i,\lambda_i\}$ be a 
discrete spectral resolution. Let
$\gamma_i(\phi):=\int_M \phi \phi_i$ be the associated Fourier 
coefficients. Then:
\begin{equation}
\beta(\phi,\rho,D,\B)(t)=
\textstyle\sum_ie^{-t\lambda_i}\gamma_i(\phi)\gamma_i(\rho).\label{AREFaA}
\end{equation}

It is convenient to introduce a formalism to consider both Dirichlet 
and Robin boundary conditions at the same time.
Suppose given a decomposition $\partial M=C_D\cup C_R$ of the boundary 
as the disjoint union of two closed (possibly
empty) sets. Let $S$ be an auxiliary endomorphism of $V|_{C_R}$ and 
let
$\phi_{;m}$ be the covariant derivative of $\phi$ with respect to the 
inward unit normal, where we use the natural
connection which is induced on $V$ by $D$ - see Section \ref{Sect2} for 
details. We define
$$\BX=\BD\oplus\BR\text{ where }\BD\phi:=\phi|_{C_D}\text{ and }
\BR\phi:=(\phi_{;m}+S\phi)|_{C_R}$$
are the pure Dirichlet and Robin operators respectively. In Section 
\ref{Sect2}, we review previous results for the boundary 
conditions $\BX$.

Transmittal and transfer boundary conditions will form the primary 
focus of this paper. Let $(M_\pm,g_\pm)$ be
smooth compact $m$ dimensional Riemannian manifolds. We assume that 
$\Sigma=\partial M_+=\partial M_-$ is a
smooth $m-1$ dimensional manifold and that the induced metrics agree, 
i.e. $g_+|_\Sigma=g_-|_\Sigma$. Let $D_\pm$ be
operators of Laplace type on vector bundles $V_\pm$ over $M_\pm$. Let
$\nu_\pm$ be the inward unit normals of $\Sigma\subset M_\pm$;
note that $\nu_+=-\nu_-$. Let $\phi:=(\phi_+,\phi_-)$ and 
$\rho:=(\rho_+,\rho_-)$.

Suppose that 
$V_+|_\Sigma=V_-|_\Sigma$ and that there is given an auxiliary
endomorphism $U$ of
$V_\Sigma:=V_\pm|_\Sigma$ serving as an impedance matching term. 
Let $\nabla^\pm$ be the natural connections
defined by the operators
$D_\pm$. Let
\begin{eqnarray}\BB\phi:&=&\{\phi_+|_\Sigma-\phi_-|_\Sigma\}\nonumber\\
&\oplus&\{(\nabla_{\nu_+}^+\phi_+)|_\Sigma
+(\nabla_{\nu_-}^-\phi_-)|_\Sigma
-U\phi_+|_\Sigma\}.\label{AREFab}\end{eqnarray}
Equivalently, $\phi$ satisfies the boundary conditions given in display 
(\ref{AREFab}) if and only if $\phi$ extends
continuously across the interface $\Sigma$ and if the normal 
derivatives match,
modulo the impedance matching term $U$. In Section \ref{Sect3}, we 
determine the invariants $\beta_n$ for $n\le3$ for
these boundary conditions, see Theorems \ref{CREFa} and \ref{CREFb} for 
details. The {\it transmittal boundary operator} $\BB=\BB(U)$ 
is of relevance in the presence of distributional sources 
\cite{BordagVassilevich99, GilkeyKirstenVassilevich01, Moss00}
as they 
have been considered, e.g., in the brane world scenario.

We shall also be studying boundary conditions which are defined
by the boundary operator $\CC=\CC(S)$:
\begin{equation}
\CC\phi:=\left\{
     \left(\begin{array}{rr}
          \nabla_{\nu_+}^++S_{++}\qquad&S_{+-}\\
          S_{-+}\qquad&\nabla_{\nu_-}^-+S_{--}\end{array}\right)
      \left(\begin{array}{l}\phi_+\\\phi_-\end{array}\right)
\right\}\bigg|_\Sigma\label{AREFac}\end{equation}
where
$$
\begin{array}{ll}
S_{++}:V_+|_\Sigma\rightarrow V_+|_\Sigma,\quad&
S_{+-}:V_-|_\Sigma\rightarrow V_+|_\Sigma,\\
S_{-+}:V_+|_\Sigma\rightarrow V_-|_\Sigma,\quad&
S_{--}:V_-|_\Sigma\rightarrow V_-|_\Sigma.
\end{array}
$$
If $S_{+-}=S_{-+}=0$, then equation (\ref{AREFac}) decouples to define 
Robin boundary
conditions. Note that we do not assume given an identification of 
$V_+|_\Sigma$ with $V_-|_\Sigma$;
in particular, we can consider the situation when $\dim V_+\ne\dim 
V_-$. In Section \ref{Sect4} we determine the
heat content invariants $\beta_n$ for $n\le 3$ for the {\it heat transfer
boundary  conditions} $\CC$, see Theorem \ref{DREFc}.

The boundary conditions defined by
equations (\ref{AREFab}) and (\ref{AREFac}) can be thought of as living 
on the
singular manifold $M:=M_+\cup_\Sigma M_-$. 
Both boundary conditions are relevant to heat transfer problems between
two media of different conductivities. Which boundary condition is to 
be 
applied depends  on the details of the surface of separation $\Sigma$
between $M_+$ and $M_-$. Let $K_+$ and $K_-$ be the 
thermal conductivities of $M_+$ and $M_-$. 
The flux of heat is continuous over the interface  
$\Sigma$,
\begin{equation}
(K_+ \nabla_{\nu_+}^+ \phi_+ + K_- \nabla_{\nu _-} ^- \phi_- ) 
\left|_\Sigma
\right. =0 .\label{jae1}
\end{equation}
If the contact between the two media $M_+$ and $M_-$ is very intimate, 
in addition
one assumes
\begin{equation}
\phi_+ \left|_\Sigma = \phi_- \right|_\Sigma , \label{jae2}
\end{equation}
and boundary conditions of the type (\ref{AREFab}) are found.
Otherwise, e.g., for surfaces pressed lightly together, in a linear 
approximation the flux of heat between $M_+$ and $M_-$ is proportional 
to
their temperature difference. In this case, equation (\ref{jae1}) 
has to be augmented by
\begin{equation}
(K_+ \nabla_{\nu_+} ^+ \phi_+)\left|_\Sigma
  = H (\phi_+ - \phi _- )\right|_\Sigma \label{jae3}
\end{equation}
where $H$ is referred to as the surface conductivity. The boundary 
conditions
(\ref{jae1}) and (\ref{jae3}) can be combined into the form of equation
(\ref{AREFac}); see, for example, the discussion in \cite{Carslaw86}.
\section{Dirichlet and Robin boundary conditions}\label{Sect2}

We begin by reviewing some of the basic invariance theory of operators 
of Laplace type. 
Let $(M,g)$ be a compact Riemannian manifold of dimension $m$. We 
suppose
the boundary $\Sigma$ of $M$ is smooth. We adopt the Einstein 
convention and sum over repeated indices. Let
$$D=-(g^{\mu\nu}\partial_\mu\partial_\nu+A^\mu\partial_\mu+B)$$
be an operator of Laplace type on $C^\infty(V)$. The operator
$D$ determines a natural connection $\nabla$ and a natural endomorphism 
$E$ such that we may express $D$ invariantly
in the form:
$$D=-\{\textstyle{Tr}(\nabla^2)+E\};$$
see \cite{Gilkey94} for details. Let
$\Gamma$ be the Christoffel symbols of the metric. We may express the 
connection
$1$ form
$\omega$ of
$\nabla$ and the endomorphism
$E$:
\begin{eqnarray}
   &&\textstyle\omega_\delta=\frac12g_{\nu\delta}
     (A^{\nu}+g^{\mu\sigma} 
    \Gamma_{\mu\sigma}{}^\nu)\text{ and}\nonumber\\
   &&{}E=B-g^{\nu\mu}(\partial_\nu{}\omega_\mu
    +\omega_\nu\omega_\mu
    -\omega_\sigma\Gamma_{\nu\mu}{}^\sigma).
    \label{BREFaa}\end{eqnarray}
Note that the connection defined by the dual operator $\tilde D$ on the dual
bundle 
$V^*$ is the associated dual connection with
connection
$1$ form given by
$-\omega^*$; furthermore the associated endomorphism is $E^*$.

We shall let Roman indices $a$, $b$, etc. range from $1$ to
$m-1$ and index a local coordinate frame for the tangent bundle of the 
boundary. Let $e_m$ be the inward unit normal
and let indices
$i$,
$j$, etc. range from $1$ to $m$ and index this augmented frame for 
$TM$.
Let
$L_{ab}$ be the second fundamental form, let 
$R_{ijkl}$ be the Riemann curvature tensor with
the sign convention that $R_{1221}=+1$ for the unit sphere in 
$\mathbb{R}^3$. Let $\Omega$ be the curvature of the
induced connection on $V$. Let `:' and `;' denote multiple covariant 
differentiation with respect to the Levi-Civita
connection of the boundary and of the interior, respectively; these two 
connections differ by the second fundamental
form. 

The invariants $\beta_n$ may be decomposed as sums
$\beta_n=\beta_n^{int}+\beta_n^{bd}$ of locally computable invariants given by
integrals over the interior and over the boundary. Let
$\tilde D$ and
$\tilde\B$ be the dual operators on
$C^\infty(V^*)$. The interior invariants are independent of the boundary 
condition and vanish if $n$ is odd. For $n\le 3$, we
have:
$$\begin{array}{ll}
\textstyle\beta_0^{int}(\rho,\phi,D,\B)=\int_M\phi\cdot\rho,
&\beta_1^{int}(\rho,\phi,D,\B)=0,\\
\textstyle\beta_2^{int}(\rho,\phi,D,\B)=-\int_MD\phi\cdot\rho,
&\beta_3^{int}(\rho,\phi,D,\B)=0.
\end{array}$$

The heat content asymptotics $\beta_n$ defined by the Dirichlet and 
Robin boundary operator $\BX$ have been studied
previously
\cite{BergDesjardinsGilkey1993, BergGilkey1999, BergGilkey2000, 
BergGilkey2001,
McAvity92, McAvity93, Savo96, Savo98, Savo98a}. There are also results 
available in the singular setting, see
for example
\cite{BergHollander99, BergSri}. We summarize the results for 
$\beta_0$, $\beta_1$, $\beta_2$, and $\beta_3$:
\begin{theorem}\label{BREFb}\ 
\begin{enumerate}
\item $\beta_0(\phi,\rho,D,\BX)=\int_{M}\phi\rho$.

\item $\beta_1(\phi,\rho,D,\BX)=-\frac2{\sqrt\pi}\int_{C_D}\phi\rho$.
\item $\beta_2(\phi,\rho,D,\BX)=-\int_{M}D\phi\cdot\rho
          +\int_{C_D}\{\frac12L_{aa}\phi\rho-\phi\rho_{;m}\}$\smallbreak
          $+\int_{C_R}\BR\phi\cdot\rho$.
\item $\beta_3(\phi,\rho,D,\BX)=
 -\frac2{\sqrt{\pi}}{\textstyle\int}_{C_D}\{-\frac23D\phi\cdot\rho 
 -\frac23 \phi\tilde D\rho+\frac13 \phi_{:a}\rho_{:a} $
\smallbreak
$ ( -\frac13E+{\textstyle\frac1{12}} L_{aa}L_{bb}
 -{\textstyle\frac16} L_{ab}L_{ab} +\textstyle\frac16 R_{am 
am})\phi\rho\}$
$+ \frac4{3\sqrt{\pi}}{\textstyle\int}_{C_R} 
\BR\phi\cdot\tilde\BR\rho$.
\end{enumerate}
\end{theorem}

\section{The boundary operator $\BB$}\label{Sect3}

We postpone for the moment the discussion of $\beta_3$. Using the chiral symmetry and the
homogeneity of the invariants, we see:

\begin{theorem} There exist universal constants so\label{CREFa}
\begin{enumerate}
\smallskip\item
$\beta_0(\phi,\rho,D,\BB)=\textstyle\int_{M_+}\phi_+\rho_++\int_{M_-}\phi_-\rho_-$.
\smallskip\item
$\beta_1(\phi,\rho,D,\BB)=\textstyle\int_\Sigma\{\coea(\phi_+\rho_++\phi_-\rho_-)+\coeb(\phi_+\rho_-+\phi_-\rho_+)\}$.
\smallskip\item
$\beta_2(\phi,\rho,D,\BB)=-\textstyle\int_{M_+}D_+\phi_+\cdot\rho_+
    -\textstyle\int_{M_-}D_-\phi_-\cdot\rho_-$
\smallbreak
$+\textstyle\int_\Sigma\big\{\coec(\phi_+\rho_+L_{aa}^++\phi_-\rho_-L_{bb}^-)
    +\coed(\phi_+\rho_+L_{aa}^-+\phi_-\rho_-L_{aa}^+)$
\smallbreak
   $+\coee(\phi_+\rho_-L_{aa}^++\phi_-\rho_+L_{aa}^-)
    +\coef(\phi_+\rho_-L_{aa}^-+\phi_-\rho_+L_{aa}^+)$
\smallbreak
   $+\coeg(\phi_{+;\nu_+}\rho_++\phi_{-;\nu_-}\rho_-)
    +\coeh(\phi_{+;\nu_+}\rho_-+\phi_{-;\nu_-}\rho_+)$
\smallbreak
   $+\coei(\phi_+\rho_{+;\nu_+}+\phi_{-}\rho_{-;\nu_-})
    +\coej(\phi_+\rho_{-;\nu_-}+\phi_{-}\rho_{+;\nu_+})$
\smallbreak
   $+\coek(\phi_+\rho_++\phi_-\rho_-)U
    +\coel(\phi_+\rho_-+\phi_-\rho_+)U\big\}.$
\item We have:
$$\begin{array}{llllll}
\textstyle\coea=-\frac1{\sqrt\pi},&\textstyle\coeb=\frac1{\sqrt\pi},&
\textstyle\coec=\frac18,&\coed=\textstyle\frac18,&

\textstyle\coee=-\frac18,&\textstyle\coef=-\frac18,
\\\textstyle\coeg=\frac12,&\textstyle\coeh=\frac12,&
\textstyle\coei=-\frac12,&\textstyle\coej=\frac12,&\textstyle\coek=-\frac14,&\textstyle\coel=-\frac14 
. 
\end{array}$$
\end{enumerate}
\end{theorem}

There are a number of functorial properties that these invariants 
satisfy. 
Suppose that the bundles $V_\pm$ are equipped
with Hermitian inner products, that the operators $D_\pm$ are formally 
self-adjoint, and that
$U$ is self-adjoint. We then have that $D$ is self-adjoint, see 
\cite{GilkeyKirstenVassilevich01}
(equation (17)) for details. We may therefore apply the relations of 
display (\ref{AREFaA}) to see:
$$\beta_n(\phi,\rho,D,\BB)=\beta_n(\rho,\phi,D,\BB).$$
More generally, if $\tilde D$ is the formal adjoint of $D$ on 
$C^\infty(V^*)$ and if $\tilde\BB$
are the dual boundary conditions, then we have
\begin{equation}
\beta_n(\phi,\rho,D,\BB)=\beta_n(\rho,\phi,\tilde 
D,\tilde\BB).\label{CREFaa}\end{equation}

The expression $-\int_MD\phi\cdot\rho$ is not symmetric in $\phi$ and 
$\rho$. We use equation
(\ref{CREFaa}) and integrate by parts to see:
\begin{equation}
\coee=\coef,\ \coeg-\coei=1,\ \coeh=\coej .\label{CREFab}
\end{equation}

Doubling the manifold yields additional information.
Let $M_0$ be a smooth Riemannian manifold with smooth boundary $\Sigma$ 
and let $D_0$ be a self-adjoint
operator of Laplace type over $M_0$.
Let $\{\tilde\phi_{D,i},\lambda_{D,i}\}$ and 
$\{\tilde\phi_{R,i},\lambda_{R,i}\}$ be the discrete
spectral resolutions for $D_0$ with Dirichlet ($D$) and Robin ($R$) 
boundary conditions over $M_0$.
Let $M^\pm:=M_0$ define the double. Extend the
$\tilde\phi_{D,i}$ to be odd and the
$\tilde\phi_{R,i}$ to be even:
$$\phi_{D,i}(x_\pm)=\pm\textstyle\frac1{\sqrt2}\tilde\phi_{D,i}(x)\text{ 
and }
  \phi_{R,i}(x_\pm)=\textstyle\frac1{\sqrt2}\tilde\phi_{R,i}(x).$$
Set $U=-2S$. It was shown in \cite{GilkeyKirstenVassilevich01} that 
$\BB\phi_{D,i}=0$ and
$\BB\phi_{R,i}=0$. Furthermore, $\{\tilde\phi_{D,i},\tilde\phi_{R,i}\}$ 
is a complete orthonormal basis
for
$L^2(V)$ which defines the spectral resolution of $D:=(D_0^+,D_0^-)$. 
Decompose
$\phi=\phi_o+\phi_e$ and $\rho=\rho_o+\rho_e$ as the sum of even and 
odd functions and let $\tilde\phi_o$,
$\tilde\phi_e$, $\tilde\rho_o$, and $\tilde\rho_e$ be the restrictions 
to $M_0=M_+$. We then have
$$\begin{array}{ll}
\tilde\phi_o=\textstyle\sum_i\gamma_{D,i}(\tilde\phi_o)\tilde\phi_{D,i},&
\phi_o=\sqrt{2}\textstyle\sum_i\gamma_{D,i}(\tilde\phi_o)\phi_{D,i},\\

\tilde\rho_o=\textstyle\sum_i\gamma_{D,i}(\tilde\rho_o)\tilde\phi_{D,i},&
\rho_o=\sqrt{2}\textstyle\sum_i\gamma_{D,i}(\tilde\rho_o)\phi_{D,i},\\
\tilde\phi_e=\textstyle\sum_i\gamma_{R,i}(\tilde\phi_e)\tilde\phi_{R,i},&
\phi_e=\sqrt{2}\textstyle\sum_i\gamma_{R,i}(\tilde\phi_e)\phi_{R,i},\\
\tilde\rho_e=\textstyle\sum_i\gamma_{R,i}(\tilde\rho_e)\tilde\phi_{R,i},&
\rho_e=\sqrt{2}\textstyle\sum_i\gamma_{R,i}(\tilde\rho_e)\phi_{R,i}.
\end{array}$$
Consequently by equation (\ref{AREFaA}),
\begin{eqnarray}
&&\beta(\phi,\rho,D,\BB)(t)=2\beta(\tilde\phi_o,\tilde\rho_o,D_0,\BD)(t)
     +2\beta(\tilde\phi_e,\tilde\rho_e,D_0,\BR)(t)\nonumber\\
&&\beta_n(\phi,\rho,D,\BB)=2\beta_n(\tilde\phi_o,\tilde\rho_o,D_0,\BD)+2\beta_n(\tilde\phi_e,\tilde\rho_e,D_0,\BR).
\label{CREFac}
\end{eqnarray}
This relation continues to hold even if $D_0$ is not self-adjoint. Thus
\begin{equation}
\begin{array}{ll}
2\coea+2\coeb=0,&2\coea-2\coeb=-\textstyle\frac4{\sqrt{\pi}},\\
2\coec+2\coed+2\coee+2\coef=0,&2\coec+2\coed-2\coee-2\coef=1,\\
2\coeg+2\coeh=2,&2\coeg-2\coeh=0,\\
2\coei+2\coej=0,&2\coei-2\coej=-2,\\
-4\coek-4\coel=2,&2\coek-2\coel=0.
\end{array}\label{CREFad}\end{equation}

Take arbitrary metrics on $M_\pm$ and let $D_\pm$ be the scalar 
Laplacian. Take $\phi=1$ and $U=0$.
Then $D\phi=0$ and $\BB\phi=0$ so $e^{-tD_\BB}\phi=\phi$. Thus
$\beta_n(1,\rho,D,\BB)=0$ for
$n\ge1$.  Take $\rho_-=0$. The terms $\rho_+$, $\rho_+L_{aa}^+$, 
$\rho_+L_{aa}^-$, $\rho_{+;\nu_+}$ can
then be specified arbitrarily. This yields:
\begin{equation}\coea+\coeb=0,\ \coec+\coef=0,\ \coed+\coee=0,\
\coei+\coej=0.\label{CREFae}\end{equation}

This allows for the determination of the multipliers $a_1,...,a_{12}$. 
However, in order to provide further checks and because it will be 
useful
later, we give
one final property. Let $N_\pm:=[0,1]$ be the interval. Let 
$M_\pm:=[0,1]\times S^1$ be the
cylinder with the metrics
$$ds^2=dr^2+e^{2f_\pm(r)}d\theta^2$$
where the real functions $f_\pm$ vanish on $\partial\{[0,1]\}$. Let 
$f_{\pm,r}:=\partial_rf_\pm$. Let
\begin{eqnarray*}
&&D_{\pm,N}:=-(\partial_r^2+f_{\pm,r}\partial_r)\text{ on }N_\pm\text{ 
and }\\
&&D_{\pm,M}=-(\partial_r^2+f_{\pm,r}\partial_r+e^{-2f_\pm}\partial_\theta^2)\text{ 
on }M_\pm.
\end{eqnarray*}
Then $D_{\pm,M}$ is the scalar Laplacian on $M_\pm$. The second 
fundamental form vanishes on $N_\pm$
while $L^\pm=-f_{\pm,r}$ is the second fundamental form on $M_\pm$. The 
connection forms defined by these two operators
differ:
$$\begin{array}{ll}
\omega^N_r=\textstyle\frac12f_r\text{ on 
}V,&\omega^N_r=-\textstyle\frac12f_r\text{ on
}V^* ,\\
\omega^M_r=0\text{ on }V,&\omega^M_r=0\text{ on }V^* . 
\end{array}$$
To compensate
for this difference, we let
$$U^N=\textstyle\frac12(f_{+,r}+f_{-,r})\text{ on }\partial N\text{ and 
}U^M=0\text{ on }\partial M.$$
The volume forms also differ:
$$\dvol^N=dr\text{ on }N_\pm\text{ and 
}\dvol^M=e^{f_\pm}drd\theta\text{ on }M_\pm.$$
We let $\phi_\pm$ and $\rho_\pm$ be constants.  We then have:
\begin{eqnarray}
&&e^{-tD_{\BB}^M}\phi=e^{-tD_{\BB}^N}\phi\text{ so}\nonumber\\
&&\beta_n(\phi,\rho,D^M,\mathcal{B}_1^M)=2\pi\beta_n(\phi,e^f\rho,D^N,\mathcal{B}_1^N).\label{CREFaex}
\end{eqnarray}
We take $f_+=f$ and $f_-=0$. 
On the cylinder, the only invariant that 
plays a role in the computation of
$\beta_2(\phi,\rho,D,\BB)$ is
$L_{aa}^+=-f_r$. On the interval, the only invariants that play a role 
are $U=\frac12f_r$, the connection $1$
form $\omega_r=\frac12f_r$ on $V$, the dual connection one form 
$-\frac12f_r$ on $V^*$, and the endomorphism
$E=-\frac14f_r^2-\frac12f_{rr}$; the interior invariants vanish as 
$$\begin{array}{llll}
D_M(\phi)=0,&D_N(\phi)=0,&
\tilde D_M(\rho)=0,&\tilde D_N(e^f\rho)=0.
\end{array}$$
Note that on $\partial N$,  
$(e^f\rho_+)_{;\nu_+}=\frac12f_r\rho_+$. Thus equation (\ref{CREFaex}) 
implies:
\begin{eqnarray*}
&&f_r(-\coec\phi_+\rho_+-\coed\phi_-\rho_--\coee\phi_+\rho_--\coef\phi_-\rho_+)\\
&=&\textstyle\frac12f_r\{\coeg\phi_+\rho_++\coeh\phi_+\rho_-+\coei\phi_+\rho_++\coej\phi_-\rho_+\\  
&&\quad+\coek(\phi_+\rho_++\phi_-\rho_-)+\coel(\phi_+\rho_-+\phi_-\rho_+)\}
\end{eqnarray*}
and consequently
\begin{equation}\label{CREFaey}
\begin{array}{ll}
-2\coec=\coeg+\coei+\coek,&-2\coed=\coek,\\
-2\coee=\coeh+\coel,& -2\coef=\coej+\coel.
\end{array}
\end{equation}

We solve the relations of displays (\ref{CREFab}), 
(\ref{CREFad}), (\ref{CREFae}), and (\ref{CREFaey}) to
complete the determination of $\beta_0$, $\beta_1$, and $\beta_2$ in 
this setting by determining the unknown
coefficients to complete the proof of Theorem \ref{CREFa}. \qedbox

Let $\omega_a:=\nabla_a^+-\nabla_a^-$ on $V$ and 
$\tilde\omega_a=-\omega_a^*$ on $V^*$; this is a chiral tensor that
changes sign if we interchange the roles of $\pm$ or of $V$ and $V^*$. 
We determine $\beta_3$ in this setting:
\goodbreak\begin{theorem}\label{CREFb}\ \begin{enumerate}
\item There exist universal constants so
$\beta_3(\phi,\rho,D,\BB)$\newline
$=\textstyle\frac1{6\sqrt{\pi}}\int_\Sigma\{
    \cfa(D_+\phi_+\cdot\rho_++\phi_+\cdot\tilde D_+\rho_+
+D_-\phi_-\cdot\rho_-+\phi_-\cdot\tilde D_-\rho_-)$\newline
$+\cfb(D_+\phi_+\cdot\rho_-+\phi_+\cdot\tilde D_-\rho_-
+D_-\phi_-\cdot\rho_++\phi_-\cdot\tilde D_+\rho_+)$\newline
$+\cfc(\omega_a\nabla_a^+\phi_+\cdot\rho_+-\omega_a\nabla_a^-\phi_-\cdot\rho_-
-\omega_a\phi_+\cdot\tilde\nabla_a^+\rho_++\omega_a\phi_-\cdot\tilde\nabla_a^-\rho_-)$\newline
$+\cfd(\omega_a\nabla_a^+\phi_+\cdot\rho_--\omega_a\nabla_a^-\phi_-\cdot\rho_+
+\omega_a\phi_+\cdot\tilde\nabla_a^-\rho_--\omega_a\phi_-\cdot\tilde\nabla_a^+\rho_+)$\newline
$+\cfe(\nabla_{\nu_+}^+\phi_+\cdot\tilde\nabla_{\nu_+}^+\rho_+
+\nabla_{\nu_-}^-\phi_-\cdot\tilde\nabla_{\nu_-}^-\rho_-)$\newline
$+\cff(\nabla_{\nu_+}^+\phi_+\cdot\tilde\nabla_{\nu_-}^-\rho_-
+\nabla_{\nu_-}^-\phi_-\cdot\tilde\nabla^+_{\nu_+}\rho_+)$\newline
$+\cfg(\nabla_a^+\phi_+\cdot\tilde\nabla_a^+\rho_+
+\nabla_a^-\phi_-\cdot\tilde\nabla_a^-\rho_-)$\newline
$+\cfh(\nabla_a^+\phi_+\cdot\tilde\nabla_a^-\rho_-
+\nabla_a^-\phi_-\cdot\tilde\nabla_a^+\rho_+)$\newline
$+\cfi 
U(\partial_{\nu_+}(\phi_+\rho_+)+\partial_{\nu_-}(\phi_-\rho_-))$\newline
$+\cfj U(\nabla_{\nu_-}^-\phi_-\cdot\rho_+
+\phi_-\cdot\tilde\nabla_{\nu_+}^+\rho_+
+\nabla_{\nu_+}^+\phi_+\cdot\rho_-+\phi_+\cdot\tilde\nabla_{\nu_-}^-\rho_-)$\newline
$+\cfk(L_{aa}^+\partial_{\nu_+}(\phi_+\rho_+)+L_{aa}^-\partial_{\nu_-}(\phi_-\rho_-))$\newline
$+\cfl(L_{aa}^-\partial_{\nu_+}(\phi_+\rho_+)+L_{aa}^+\partial_{\nu_-}(\phi_-\rho_-))$\newline
$+\cfm(L_{aa}^+(\nabla_{\nu_+}^+\phi_+\cdot\rho_-
+\phi_-\tilde\nabla_{\nu_+}^+\rho_+)
+L_{aa}^-(\nabla_{\nu_-}^-\phi_-\cdot\rho_++\phi_+\tilde\nabla_{\nu_-}^-\rho_-))$\newline
$+\cfmx(L_{aa}^-(\nabla_{\nu_+}^+\phi_+\cdot\rho_-+\phi_-\tilde\nabla_{\nu_+}^+\rho_+)
+L_{aa}^+(\nabla_{\nu_-}^-\phi_-\cdot\rho_++\phi_+\tilde\nabla_{\nu_-}^-\rho_-))$\newline
$+\cfn\omega_a\omega_a(\phi_+\rho_++\phi_-\rho_-)
+\cfo \omega_a\omega_a(\phi_+\rho_-+\phi_-\rho_+)$\newline
$+\cfp(L_{aa}^+L_{bb}^+\phi_+\rho_++L_{aa}^-L_{bb}^-\phi_-\rho_-)
   +\cfq L_{aa}^+L_{bb}^-(\phi_+\rho_++\phi_-\rho_-)$\newline
$+\cfr(L_{aa}^-L_{bb}^-\phi_+\rho_++L_{aa}^+L_{bb}^+\phi_-\rho_-)
  +\cfs L_{aa}^+L_{bb}^-(\phi_+\rho_-+\phi_-\rho_+)$\newline
$+\cft(L_{aa}^+L_{bb}^++L_{aa}^-L_{bb}^-)(\phi_+\rho_-+\phi_-\rho_+)$\newline
$+\cfu(L_{ab}^+L_{ab}^+\phi_+\rho_++L_{ab}^-L_{ab}^-\phi_-\rho_-)
   +\cfv L_{ab}^+L_{ab}^-(\phi_+\rho_++\phi_-\rho_-)$\newline
$+\cfw(L_{ab}^-L_{ab}^-\phi_+\rho_++L_{ab}^+L_{ab}^+\phi_-\rho_-)
  +\cfx L_{ab}^+L_{ab}^-(\phi_+\rho_-+\phi_-\rho_+)$\newline
$+\cfy(L_{ab}^+L_{ab}^++L_{ab}^-L_{ab}^-)(\phi_+\rho_-+\phi_-\rho_+)$\newline
$+\cfz U(L_{aa}^+\phi_+\rho_++L_{aa}^-\phi_-\rho_-)
+\cfaa U(L_{aa}^-\phi_+\rho_++L_{aa}^+\phi_-\rho_-)$\newline
$+\cfab U(L_{aa}^++L_{aa}^-)(\phi_+\rho_-+\phi_-\rho_+)$\newline
$+\cfac U^2(\phi_+\rho_++\phi_-\rho_-)+\cfad 
U^2(\phi_+\rho_-+\phi_-\rho_+)$\newline
$+\cfae(E_+\phi_+\rho_++E_-\phi_-\rho_-)+\cfaf(E_-\phi_+\rho_++E_+\phi_-\rho_-)$\newline
$+\cfag(E_++E_-)(\phi_+\rho_-+\phi_-\rho_+)
 +\cfah(R_{ijji}^+\phi_+\rho_++R_{ijji}^-\phi_-\rho_-)$\newline
$+\cfai(R_{ijji}^-\phi_+\rho_++R_{ijji}^+\phi_-\rho_-)
  +\cfaj(R_{ijji}^++R_{ijji}^-)(\phi_+\rho_-+\phi_-\rho_+)$\newline
$+\cfak(R_{amma}^+\phi_+\rho_++R_{amma}^-\phi_-\rho_-)$\newline
  $+\cfal(R_{amma}^-\phi_+\rho_++R_{amma}^+\phi_-\rho_-)$\newline
$+\cfam(R_{amma}^++R_{amma}^-)(\phi_+\rho_-+\phi_-\rho_+)$.
\item  $\cfa =\phantom{-}4,\quad   \cfb =-4,\quad  \cfc =-1,\quad  \cfd =-1,\quad  \cfe 
=\phantom{-}4,$\newline$
 \cff =\phantom{-}4,\quad   \cfg =-2,\quad  \cfh =\phantom{-}2,\quad   \cfi =-2,\quad  \cfj 
=-2,$\newline$
 \cfk =-1,\quad  \cfl =\phantom{-}1,\quad   \cfm =\phantom{-}1,\quad  \cfmx =-1,\quad  \cfn 
=\phantom{-}1,$\newline$
  \cfo =\phantom{-}0,\quad  \cfp =\phantom{-}0,\quad   \cfq =-\textstyle\frac12,\quad  
\cfr =\phantom{-}0,\quad  \cfs
=\phantom{-}\textstyle\frac12,$\newline$
 \cft  =\phantom{-}0,\quad  \cfu =\phantom{-}\textstyle\frac12,\quad  \cfv 
=\phantom{-}0,\quad  \cfw =\phantom{-}\textstyle\frac12,\quad \cfx
=\phantom{-}0,$\newline$
 \cfy =-\textstyle\frac12,\quad  \cfz =\phantom{-}1,\quad  \cfaa =-1,\quad  \cfab 
=\phantom{-}0,\quad  \cfac =\phantom{-}1,$\newline$
 \cfad =\phantom{-}1,\quad  \cfae =\phantom{-}1,\quad  \cfaf =\phantom{-}1,\quad   
\cfag =-1,\quad  \cfah =\phantom{-}0,$\newline$
 \cfai =\phantom{-}0,\quad  \cfaj =\phantom{-}0,\quad  \cfak 
=\phantom{-}\textstyle\frac12,\quad  \cfal =\phantom{-}\textstyle\frac12,\quad 
\cfam =-\textstyle\frac12.$\end{enumerate}
\end{theorem}

We use the
relations of equation (\ref{CREFaa}) to simplify the format at the 
outset and derive (1). We shall use the
functorial properties involved to determine the unknown coefficients 
and prove (2).

We apply Theorem
\ref{BREFb} and equation (\ref{CREFac}) to see:
$$\begin{array}{ll}
2\cfa+2\cfb=0,&2\cfa-2\cfb=16,\\
%2\cfc+2\cfd=0,&2\cfc-2\cfd=0,\\
2\cfe+2\cff=16,&2\cfe-2\cff=0,\\
2\cfg+2\cfh=0,&2\cfg-2\cfh=-8,\\
-4\cfi-4\cfj=16,&2\cfi-2\cfj=0,\\
\cfk+\cfl+\cfm+\cfmx=0,&2\cfk+2\cfl-2\cfm-2\cfmx=0,\\
\cfp+\cfq+\cfr+\cfs+2\cft=0,&\cfp+\cfq+\cfr-\cfs-2\cft=-1,\\
\cfu+\cfv+\cfw+\cfx+2\cfy=0,&\cfu+\cfv+\cfw-\cfx-2\cfy=2,\\
2\cfz+2\cfaa+4\cfab=0,&2\cfz+2\cfaa-4\cfab=0,\\
8\cfac+8\cfad=16,&2\cfac-2\cfad=0,\\
2\cfae+2\cfaf+4\cfag=0,&2\cfae+2\cfaf-4\cfag=8,\\
2\cfah+2\cfai+4\cfaj=0,&2\cfah+2\cfai-4\cfaj=0,\\
2\cfak+2\cfal+4\cfam=0,&2\cfak+2\cfal-4\cfam=4.
\end{array}$$

Take arbitrary metrics on $M_\pm$ and let $D_\pm$ be the scalar 
Laplacian. Take $\phi=1$ and $U=0$.
Then $\phi$ satisfies transmittal boundary conditions. Thus
$\beta_n(1,\rho,D,\BB)=0$ for
$n\ge1$.  Take $\rho_-=0$. This yields:
$$\begin{array}{lll}
\cfk+\cfm=0,&\cfl+\cfmx=0,\\
\cfp+\cft=0,&\cfr+\cft=0,&\cfq+\cfs=0,\\
\cfu+\cfy=0,&\cfw+\cfy=0,&\cfv+\cfx=0,\\
\cfae+\cfag=0,&\cfaf+\cfag=0,&\cfah+\cfaj=0,\\
\cfai+\cfaj=0,&\cfak+\cfam=0,&\cfal+\cfam=0.
\end{array}$$

Let $D_\pm(\varepsilon)=D_\pm-\varepsilon$. Then $\tilde 
D_\pm(\varepsilon)=\tilde D_\pm-\varepsilon$
and $E_\pm(\varepsilon)=E_\pm+\varepsilon$. As 
$e^{-tD_\BB(\varepsilon)}=e^{t\varepsilon}e^{-tD_\BB}$,
$\beta(\phi,\rho,D(\varepsilon),\BB)(t)=e^{t\varepsilon}\beta(\phi,\rho,D,\BB)(t)$
and hence $\partial_\varepsilon\beta_n|_{\varepsilon=0}=\beta_{n-2}$. 
Thus studying the coefficients of the terms
$\{\phi_+\rho_+,\phi_+\rho_-\}$ leads to the relations:
\begin{eqnarray*}
&&\textstyle\frac1{6\sqrt{\pi}}\{-2\cfa+\cfae+\cfaf\}=\coea=-\frac1{\sqrt\pi},\nonumber\\
&&\textstyle\frac1{6\sqrt{\pi}}\{-2\cfb+2\cfag\}=\coeb=\frac1{\sqrt\pi}. 
\end{eqnarray*}

We use separation of variables to generate additional relationships 
among the coefficients. First, we study flat
metrics. Let $(r,\theta)$ be the usual parameters on $M_{\pm}:=[0,1]\times S^1$. Let
$$D_\pm:=-(\partial_r^2+\partial_\theta^2+2\varepsilon_\pm\partial_\theta)$$
where $\varepsilon_\pm=\varepsilon_\pm(\theta)$.
Let $N_\pm:=[0,1]$ and let $\bar D_\pm:=-\partial_r^2$. Let $\phi_\pm$ 
and
$\rho_\pm$ be constant. Let $U_0$ be constant. Separation of variables 
and an application of equation 
(\ref{CREFac}) and Theorem \ref{BREFb} then yields:
\begin{eqnarray*}
&&e^{-tD_\BB}\phi=e^{-t\bar D_\pm}\phi\text{ so}\\
&&\beta_3(\phi,\rho,D,\BB)_M=2\pi\beta_3(\phi,\rho,\bar D,\BB)_{N}\\
&&\qquad=4\pi\beta_3(\tilde\phi_e,\tilde\rho_e,\bar D,\BR)_{N_+}
+4\pi\beta_3(\tilde\phi_o,\tilde\rho_o,\bar
D,\BD)_{N_+}=0.\end{eqnarray*}
As $\nabla_\theta\phi_\pm=\varepsilon_\pm\phi_\pm$, 
$\nabla_\theta\rho_\pm=-\varepsilon_\pm\rho_\pm$, and
$E_\pm=-\varepsilon_\pm^2-\partial_\theta\varepsilon_\pm$, we have:
\medbreak\qquad
$0=\cfc(\varepsilon_+-\varepsilon_-)(2\varepsilon_+\phi_+\rho_+-2\varepsilon_-\phi_-\rho_-)$
\par\qquad\quad$
+\cfd(\varepsilon_+-\varepsilon_-)^2(\phi_+\rho_-+\phi_-\rho_+)$
\par\qquad\quad$+\cfg(-\varepsilon_+^2\phi_+\rho_+-\varepsilon_-^2\phi_-\rho_-)
+\cfh\varepsilon_+\varepsilon_-(-\phi_+\rho_--\phi_-\rho_+)$
\par\qquad\quad$+\cfn(\varepsilon_+-\varepsilon_-)^2(\phi_+\rho_++\phi_-\rho_-)+\cfo
(\varepsilon_+-\varepsilon_-)^2(\phi_+\rho_-+\phi_-\rho_+)$
\par\qquad\quad$+\cfae(-\varepsilon_+^2\phi_+\rho_+-\varepsilon_-^2\phi_-\rho_-)
+\cfaf(-\varepsilon_-^2\phi_+\rho_+-\varepsilon_+^2\phi_-\rho_-)$
\par\qquad\quad$+\cfag(-\varepsilon_+^2-\varepsilon_-^2)(\phi_+\rho_-+\phi_-\rho_+).$\medbreak\noindent
The terms $\{\varepsilon_+^2\phi_+\rho_+,\ \varepsilon_-^2\phi_+\rho_+,\
\varepsilon_+\varepsilon_-\phi_+\rho_+,\varepsilon_+^2\phi_+\rho_-,
\varepsilon_+\varepsilon_-\phi_+\rho_-\}$ are then studied to conclude that:
$$\begin{array}{ll}
0=2\cfc-\cfg+\cfn-\cfae,&
0=\cfn-\cfaf,\\
0=-2\cfc-2\cfn,&
0=\cfd+\cfo-\cfag,\\
0=-2\cfd-\cfh-2\cfo. & \end{array}$$
We can also get information from the divergence terms. We now let $\phi=(1,1)$ and
$\rho=(\rho_+(\theta),0)$. 
We work modulo $O(\varepsilon^2)$ 
and use the fact that $\cfa+\cfb=0$ to show:
\begin{eqnarray*}
&&0=\textstyle\int_\Sigma\big\{(\partial_\theta\rho_+)
\{-\cfc(\varepsilon_+-\varepsilon_-)  
-\cfd(\varepsilon_+-\varepsilon_-)+\cfg\varepsilon_++\cfh\varepsilon_-\}\\
&&\qquad+\rho_+\{-\cfae\partial_\theta\varepsilon_+-\cfaf\partial_\theta\varepsilon_-
  -\cfag\partial_\theta(\varepsilon_+
+\varepsilon_-)\}\big\}\\
&&-\cfc-\cfd+\cfg+\cfae+\cfag=0,\nonumber\\
&&\phantom{-}\cfc+\cfd+\cfh+\cfaf+\cfag=0.
\end{eqnarray*}

Next, let $N_\pm:=[0,1]$ and $M_\pm:=[0,1]\times 
S^1\times S^1$ have the metrics:
$$ds^2=dr^2+e^{2f_{1,\pm}(r)}d\theta_1^2+e^{2f_{2,\pm}(r)}d\theta_2^2$$
where $f_{i,\pm}$ vanishes on the boundary. Let 
$f_{i,\pm,r}:=\partial_rf_{i,\pm}$ and let
\begin{eqnarray*}
&&D_\pm^N:=-\{\partial_r^2+\textstyle\sum_if_{i,\pm,r}\partial_r\},\\
&&D_\pm^M:=-\{\partial_r^2+\textstyle\sum_i(f_{i,\pm,r}\partial_r+e^{-2f_{i,\pm}}\partial_\theta^2)\}. 
\end{eqnarray*}
We set $U_M=U_0$ constant. As the connection forms defined by $D^N$ and 
$D^M$ are different, we set
$U_N=U_0+\textstyle\frac12\sum_i(f_{i,+,r}+f_{i,-,r})$. Let $\phi_\pm$ 
and $\rho_\pm$ be constant.
The argument used to establish equation (\ref{CREFaex}) then 
generalizes immediately to yield:
\begin{equation}\label{CREFbea}
\beta_3(\phi,\rho,D_M,\mathcal{B}_{1,M})=(2\pi)^2\beta_3(\phi,e^{\sum_if_i}\rho,D_N,\mathcal{B}_{1,N}).
\end{equation}
We compute on $\partial M$:
\begin{eqnarray*}
\Gamma_{mab}^\pm&=&\textstyle\frac12\partial_mg_{ab}^\pm=e^{2f_{a,\pm}}f_{a,\pm,r}\delta_{ab},\\
\Gamma_{abm}^\pm&=&-\textstyle\frac12\partial_mg_{ab}^\pm=-e^{2f_{a,\pm}}f_{a,\pm,r}\delta_{ab},\\
\Gamma_{amb}^\pm&=&-\Gamma_{abm}^\pm=e^{2f_{a,\pm}}f_{a,\pm,r}\delta_{ab},\\
\Gamma_{am}^\pm{}^b&=&g^{bb}\Gamma_{amb}^\pm=f_{a,\pm,r}\delta_{ab},\\
L_{ab}^\pm&=&(\nabla_a\partial_b,\partial_m)=\Gamma_{abm}^\pm=-e^{2f_{a,\pm}}f_{a,\pm,r}\delta_{ab} 
\, ,\\
R_{amma}^\pm&=&((\nabla_a^\pm\nabla_m^\pm-\nabla_m^\pm\nabla_a^\pm),\partial_m,\partial_a)
=-(\nabla_m^\pm\Gamma_{am}^\pm{}^b\partial_b,\partial_m)\\
&=&-(\nabla_m^\pm 
f_{a,\pm,r}\partial_a,\partial_a)=-f_{a,\pm,rr}-f_{a,\pm,r}^2.
\end{eqnarray*}
(We do not compute $R_{ijji}^\pm$ as we have shown
$\cfah=\cfai=\cfaj=0$ so these terms do not appear).
Let $f_\pm=\textstyle\sum_if_{i,\pm}$. We compute on $\partial N$:\goodbreak
\begin{eqnarray*}
&&\begin{array}{ll}\omega_r^\pm=\textstyle\frac12f_{\pm,r},&
\tilde\omega_r^\pm=-\textstyle\frac12f_{\pm,r},\\
U_N=U_0 + \textstyle\frac12(f_{+,r}+f_{-,r}),\qquad&
E^\pm_N=-\textstyle\frac12f_{\pm,rr}-\textstyle\frac14f_{\pm,r}^2
\, ,\\
\nabla_\nu^\pm\rho_\pm=\frac12f_\pm\phi_\pm,&\nabla_\nu^\pm(e^{f_\pm}\rho)=\frac12f_\pm\rho_\pm,\\
D_\pm^N\phi_\pm=0,&\tilde D_\pm^N\{e^{f_\pm}\rho_\pm\}=0.
\end{array}\end{eqnarray*}
We use equation (\ref{CREFbea}) to derive the following equations from 
the
coefficients of the indicated monomials:\goodbreak
$$\begin{array}{ll}-\cfak=-\textstyle\frac12\cfae&(f_{1,+,rr}\phi_+\rho_+)\\
-\cfal=-\textstyle\frac12\cfaf&(f_{1,-,rr}\phi_+\rho_+)\\
-\cfam=-\textstyle\frac12\cfag&(f_{1,+,rr}\phi_+\rho_-)\\
\cfp+\cfu-\cfak\textstyle=\frac14\cfe+\frac12\cfi+\frac14\cfac-\textstyle\frac14\cfae\quad&
      (f_{1,+,r}f_{1,+,r}\phi_+\rho_+)\\
\textstyle\cfr+\cfw-\cfal=\frac14\cfac-\frac14\cfaf
&(f_{1,-,r}f_{1,-,r}\phi_+\rho_+)\\
\textstyle\cfq+\cfv=\frac12\cfi+\frac12\cfac&(f_{1,+,r}f_{1,-,r}\phi_+\rho_+)\\
\textstyle2\cfp=\frac12\cfe+\cfi+\frac12\cfac-\frac12\cfae&(f_{1,+,r}f_{2,+,r}\phi_+\rho_+)\\
\textstyle2\cfr=\frac12\cfac-\frac12\cfaf&(f_{1,-,r}f_{2,-,r}\phi_+\rho_+)\\
\textstyle\cfq=\frac12\cfi+\frac12\cfac&(f_{1,+,r}f_{2,-,r}\phi_+\rho_+)\\
\textstyle\cft+\cfy-\cfam=\frac14\cfj+\frac14\cfad-\frac14\cfag&(f_{1,+,r}f_{1,+,r}\phi_+\rho_-)\\
\textstyle\cfs+\cfx=\frac14\cff+\frac12\cfj+\frac12\cfad&(f_{1,+,r}f_{1,-,r}\phi_+\rho_-)\\
\textstyle2\cft=\frac12\cfj+\frac12\cfad-\frac12\cfag&(f_{1,+,r}f_{2,+,r}\phi_+\rho_-)\\
\textstyle\cfs=\frac14\cff+\frac12\cfj+\frac12\cfad&(f_{1,+,r}f_{2,-,r}\phi_+\rho_-)\\
\textstyle-\cfz=\cfi+\cfac&(U_0f_{1,+,r}\phi_+\rho_+)\\
\textstyle-\cfaa=\cfac&(U_0f_{1,-,r}\phi_+\rho_+)\\
\textstyle-\cfab=\frac12\cfj+\cfad&(U_0f_{1,+,r}\phi_+\rho_-)
\end{array}$$

We now let $\rho=\rho(r)$. Equation (\ref{CREFbea}) continues to hold 
and yields:
$$
\begin{array}{ll}
-\cfl=\frac12\cfi&(f_{1,-,r}\phi_+\partial_{\nu_+}\rho_+)\\
-\cfmx=\frac12\cff+\frac12\cfj
\quad\qquad\qquad\qquad\qquad\qquad&(f_{1,+,r}\phi_+\partial_{\nu_-}\rho_-)
\end{array}$$
We combine the relations given above to complete the determination of $\beta_3$ by 
determining the constants $\cfa-\cfam$ and complete the proof
of Theorem \ref{CREFb}.
\qedbox

\section{The boundary condition $\CC$}\label{Sect4}

Again, we begin our discussion by studying $\beta_0$, $\beta_1$, and 
$\beta_2$.
Note that terms such as $\phi_+ \rho_-$ cannot 
occur since we do not assume an identification of 
$V_+\left|_\Sigma\right.$
with $V_- \left| _\Sigma\right.$. 
\begin{lemma} There exist universal constants so\label{DREFa}
\begin{enumerate}
\smallskip\item
$\beta_0(\phi,\rho,D,\BB)=\textstyle\int_{M_+}\phi_+\rho_++\int_{M_-}\phi_-\rho_-$.
\smallskip\item
$\beta_1(\phi,\rho,D,\CC)=\textstyle\int_\Sigma\doea(\phi_+\rho_++\phi_-\rho_-)$.
\smallskip\item
$\beta_2(\phi,\rho,D,\CC)=-\textstyle\int_{M_+}D_+\phi\cdot\rho_+
    -\textstyle\int_{M_-}D_-\phi_-\cdot\rho_-$
\smallbreak
$+\textstyle\int_\Sigma\{\doeb(\phi_+\rho_+L_{aa}^++\phi_-\rho_-L_{bb}^-)
    +\doec(\phi_+\rho_+L_{aa}^-+\phi_-\rho_-L_{aa}^+)$
\smallbreak
   $+\doed(\phi_{+;\nu_+}\rho_++\phi_{-;\nu_-}\rho_-)
    +\doee(\phi_+\rho_{+;\nu_+}+\phi_{-}\rho_{-;\nu_-})$
\smallbreak
   $+\doef(S_{++}\phi_+\cdot\rho_++S_{--}\phi_-\cdot\rho_-)
    +\doeg(S_{+-}\phi_-\cdot\rho_++S_{-+}\phi_+\cdot\rho_-)\}$.
\end{enumerate}
\end{lemma}

Taking $S_{+-}=0$ and $S_{-+}=0$ 
forces the boundary conditions given in equation (\ref{AREFac}) to
decouple and defines Robin boundary conditions on $M_+$ and on $M_-$ 
separately. We use Theorem
\ref{BREFb} to see:
\begin{equation}
\begin{array}{llllll}
    \doea=0,&\doeb=0,&\doec=0,&
    \doed=1,&\doee=0,&\doef=1.\end{array}\label{DREFaa}\end{equation}
 We use an argument similar to that used to establish display (\ref{CREFae}) to determine 
$\doeg$. Let $D_\pm$ be the scalar
Laplacians on manifolds $M_\pm$. Let $\phi_+=\phi_-=1$, let 
$S_{++}=S_{--}=1$, and let
$S_{+-}=S_{-+}=-1$. Then $\CC\phi=0$ and $D\phi=0$ so
$\beta_n=0$ for $n\ge0$. Thus:
\begin{equation}\doef-\doeg=0.\label{DREFab}\end{equation}

In view of the remarks noted above, we see that:
\begin{lemma}\label{DREFb}
There exist  universal constants so
\medbreak\quad
$\beta_3(\phi,\rho,D,\CC)=\frac4{3\sqrt\pi}\textstyle\int_\Sigma\CC\phi\cdot\tilde\CC\rho$
\medbreak\qquad
$+\dfa(S_{+-}S_{-+}\phi_+\cdot\rho_++S_{-+}S_{+-}\phi_-\cdot\rho_-)$
\medbreak\qquad
$+\dfb(S_{--}S_{-+}\phi_+\cdot\rho_-+S_{++}S_{+-}\phi_-\cdot\rho_+)$
\medbreak\qquad
$+\dfc(S_{-+}S_{++}\phi_+\cdot\rho_-+S_{+-}S_{--}\phi_-\cdot\rho_+)$
\medbreak\qquad
$+\dfd(S_{-+}\phi_{+;\nu_+}\cdot\rho_-+S_{+-}\phi_{-;\nu_-}\cdot\rho_+)$
\medbreak\qquad
$+\dfe(S_{-+}\phi_+\cdot\rho_{-;\nu_-}+S_{+-}\phi_-\cdot\rho_{+;\nu_+})$
\medbreak\qquad
$+\dff 
(L_{aa}^+S_{-+}\phi_+\cdot\rho_-+L_{aa}^-S_{+-}\phi_-\cdot\rho_+)$
\medbreak\qquad
$+\dfg 
(L_{aa}^-S_{-+}\phi_+\cdot\rho_-+L_{aa}^+S_{+-}\phi_-\cdot\rho_+)$.
\end{lemma}

Let $D_\pm$ be the scalar
Laplacians on manifolds $M_\pm$. Let $\phi_+=1$, $\phi_-=1$, let 
$S_{++}=a$, 
$S_{+-}=-a$, $S_{--}=b$, and $S_{-+}=-b$. Then $\phi$ satisfies 
transmittal boundary conditions and is harmonic so
$\beta_n=0$ for $n\ge0$. Consequently taking $\rho_-=0$ yields the 
equations:
\begin{equation}
\dfa-\dfc=0,\ \dfb=\dfe=\dff=\dfg=0.
\label{DREFba}
\end{equation}
We work with $m=1$ and $D_\pm=-\partial_x^2$. Suppose that 
$\phi_\pm=a_\pm x+b_\pm$ is such that $\phi$ satisfies
$\CC\phi=0$, i.e.
\begin{equation}\varepsilon a_++S_{++}b_++S_{+-}b_-=0\text{ and }
\varepsilon a_-+S_{--}b_-+S_{-+}b_+=0\label{DREFbbx}\end{equation}
where $\varepsilon(0)=+1$ and $\varepsilon(1)=-1$. We choose $S_*$ and $b_*$ arbitrarily and use equation
(\ref{DREFbbx}) to determine
$a_*$. Let $\rho_-=0$. Since $\beta_n=0$ for $n>0$,
\begin{eqnarray}
&&\dfa S_{+-}S_{-+}b_++\dfc S_{+-}S_{--}b_-+\dfd\varepsilon S_{+-}a_-=0,\nonumber\\
&&(\dfa-\dfd)S_{+-}S_{-+}b_++(\dfc-\dfd)S_{+-}S_{--}b_-=0\text{ so }\nonumber\\
&&\dfa=\dfd, \quad\text{ and } \quad\dfc=\dfd 
.\label{DREFbb}\end{eqnarray}

We double the manifold to complete our determination. Suppose given an 
operator $D_0$ of Laplace type on a manifold
$M_0$ with boundary $\Sigma$. Let an initial condition $\phi_0$ be 
given and let $u_0$ solve equation (\ref{AREFaa}) 
with the boundary operator $\BR$. Let
$M_\pm:=M_0$ and
$D_\pm:=D_0$. Then $u_\pm:=u_0$ and $\phi_\pm:=\phi_0$ solves equation 
(\ref{AREFaa}) with
the boundary operator
$\CC$ defined by equation (\ref{AREFac}) with
$S_{++}=S_{--}=0$ and $S_{+-}=S_{-+}=S$.  Thus
$$\beta_n(\phi_0,\rho_++\rho_-,D_0,\BR)=\beta_n(\phi,\rho,D,\CC).$$
We may now conclude $\dfa=\dfd=0$. This proves

\begin{theorem}\ 
\label{DREFc}\begin{enumerate}
\item $\beta_0(\phi,\rho,D,\CC)=\int_{M}\phi\rho$.
\item $\beta_1(\phi,\rho,D,\CC)=0$.
\item $\beta_2(\phi,\rho,D,\CC)=-\int_{M}D\phi\cdot\rho
          +\int_\Sigma\CC\phi\cdot\rho$.
\item 
$\beta_3(\phi,\rho,D,\CC)=\frac4{3\sqrt{\pi}}{\textstyle\int}_\Sigma\CC\phi\cdot\tilde\CC\rho$.
\end{enumerate}
\end{theorem} 
\noindent {\bf Acknowledgements:} \\[5pt]
It is a pleasant task to thank Michiel van den Berg for helpful conversations
concerning various matters. Research of P. Gilkey was partially supported by the
NSF (USA) and MPI 
(Leipzig). Research of K. Kirsten was partially supported by the MPI (Leipzig).

\medbreak\noindent\hrule\medbreak\noindent
PG: Department of Mathematics, University of
Oregon, Eugene OR 97403 USA. EMAIL: gilkey@darkwing.uoregon.edu
\medbreak\noindent
KK: Max Planck Institute for Mathematics in the Sciences, Inselstrasse 22-26,
04103 Leipzig, Germany. EMAIL: klaus.kirsten@mis.mpg.de 
\end{document}